\documentclass[pteplogo]{ptephy}
\usepackage{graphics}
\usepackage{amsmath}
\begin{document}
\newcommand{\op}{\boldsymbol}

\title{Three-Slit Interference: A Duality Relation}

\author{\name{Mohd Asad Siddiqui}{} and \name{Tabish Qureshi}{}}
\address{\affil{}{Centre for Theoretical Physics, Jamia Millia Islamia, New Delhi, India.}
\email{tabish@ctp-jamia.res.in}}


\begin{abstract}

The issue of interference and which-way information is addressed in the 
context of 3-slit interference experiments. A new path
distinguishability ${\mathcal D_Q}$ is introduced, based on Unambiguous
Quantum State Discrimination (UQSD).
An inequality connecting the
interference visibility and path distinguishability,
${\mathcal V} + {2{\mathcal D_Q}\over 3- {\mathcal D_Q}} \le 1$, is derived which
puts a bound on how much fringe visibility and which-way information can be
simultaneously obtained. It is argued that this bound is tight.
For 2-slit interference, we derive a new duality relation which reduces
to Englert's duality relation and Greenberger-Yasin's duality relation,
in different limits.
\end{abstract}



\maketitle

\section{Introduction}

The two-slit interference experiment with particles has become a cornerstone
for studying wave-particle duality. So fundamental is the way in which
the two-slit experiment captures the essence of quantum theory, that
Feynman ventured to state that it is a
phenomenon ``which has in it the heart of quantum mechanics; in reality it
contains the {\em only} mystery" of the theory \cite{feynman}.
That radiation and massive particles can
exhibit both wave nature and particle nature in different experiment, had
become quite clear in the early days of quantum mechanics. However, Niels Bohr
emphasized that wave-nature, characterized by two-slit interference, and
the particle-nature, characterized by the knowledge of which slit the
particle passed through, are mutually exclusive \cite{bohr}.
In doing this he raised this concept to the level of a new fundamental
principle. 

Much later, this principle was made quantitatively precise by deriving a
bound on the extent to which the two natures could be observed simultaneously,
by Greenberger and Yasin \cite{greenberger} and later by 
Englert \cite{englert}. Greenberger and Yasin characterised the particle
nature by the ability to correctly predict which slit the particle passed
through \cite{greenberger}. This predictability was based only on the
initial state of the particle, and not on any measurement on it.
Englert characterised the particle nature by the ability to distinguish
between the two paths of the particle, by an actual measurement \cite{englert}.
He introduced a quantity ${\mathcal D}$ for this purpose, which took values
between 0 and 1. The wave nature was characterised by the visibility of the
interference, given by ${\mathcal V}$. The relation putting a bound on the
path distinguishability and fringe visibility is given by \cite{englert}
\begin{equation}
{\mathcal V}^2 + {\mathcal D}^2 \le 1.
\label{egy}
\end{equation}
Thus one can see that ${\mathcal D}$ and ${\mathcal V}$, which can take
values between 0 and 1, are dependent on each other. A full which-way
information (${\mathcal D}=1$) would surely wash out the interference
(${\mathcal V}=0$). Eqn. (\ref{egy}) can be thought to be a quantitative
statement of Bohr's complementarity principle. It smoothly interpolates
between the two extreme scenarios discussed by Bohr, namely, full which-way
information and no which-way information.

\begin{figure}
\centerline{\resizebox{13cm}{!}{\includegraphics{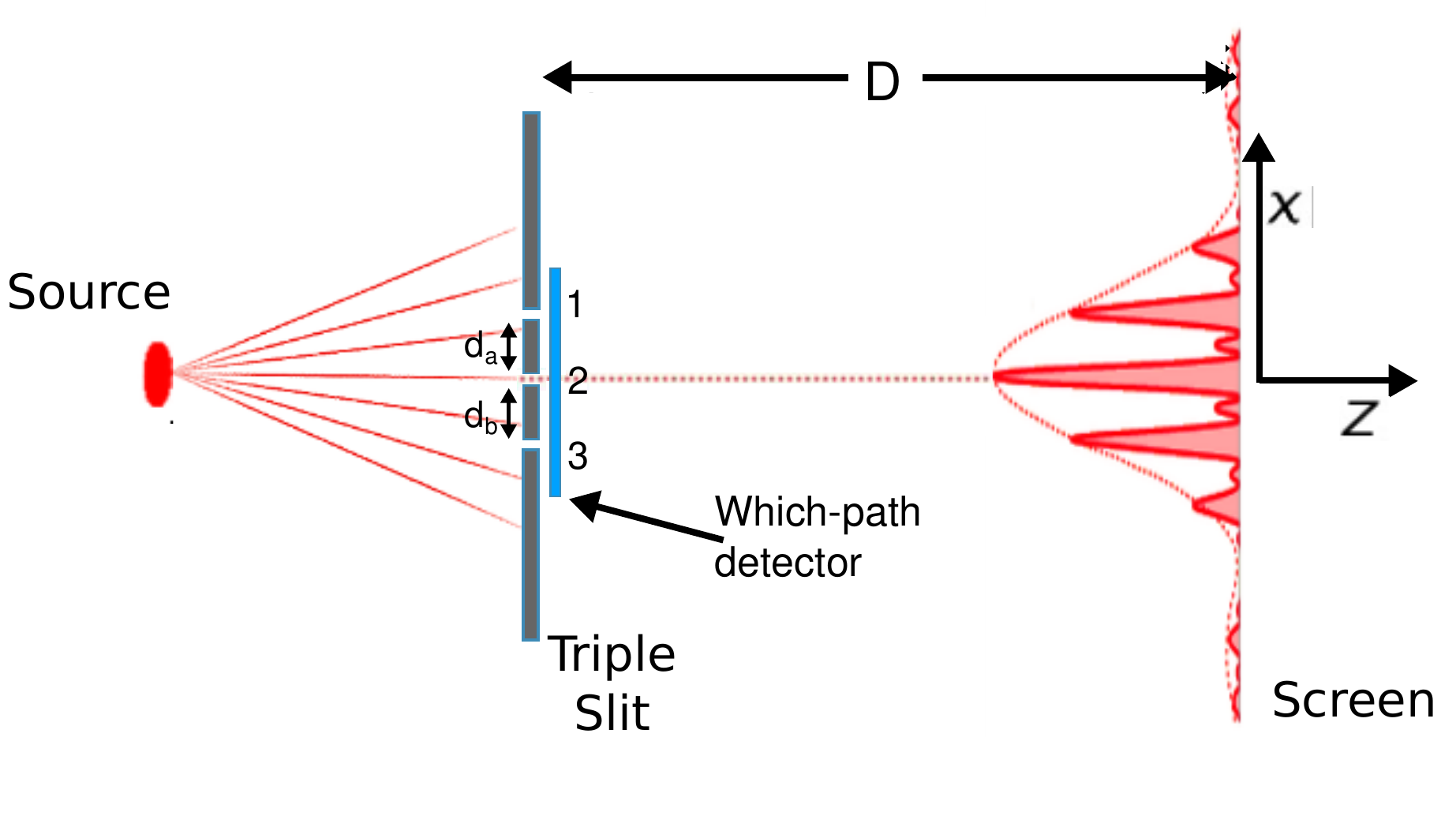}}}
\caption{A schematic diagram of the 3-slit interference experiment,
with a quantum which-path detector. 
}
\label{trislit}
\end{figure}

A dramatic manifestation of Bohr's complementarity principle has been 
demonstrated in the so-called quantum eraser. Here, ``erasing" the which-way
information after the particle has passed through the slits, allows
one to recover the lost interference fringes \cite{marcelo}.
Wave-particle duality has also been connected to various other phenomena.
Connection of wave-particle duality with uncertainty relations has
been investigated \cite{rempe,tqrv,coles,tanimura}.
An interesting complementarity between entanglement and the visibility
of interference has been demonstrated for two-path interference \cite{hosoya}. 
That the entanglement between the particle and which-path detector is
what affects the visibility of interference, is also the view we take in
this investigation.

Various aspects of complementarity were also explored by Jaeger, 
Shimony and Vaidman \cite{jaeger} where they also explored the case
where the particle can follow multiple path, and not just two, before
interfering. Bohr's principle
of complementarity should surely apply to multi-slit experiments too. However,
one might wonder if one can find a quantitatively precise statement
of it for multi-slit experiments. Various attempts have been made to
formulate a quantitatively precise statement of complementarity, some
kind of a duality relation, for the case of
multibeam interferometers \cite{durr,bimonte,bimonte1,englertmb,zawisky}. However, the issue
is still not satisfactorily resolved. Beyond the well studied two-slit
experiment, the simplest multi-beam case is the 3-slit interference experiment.
Englert's duality relation was derived only in the 
context of 2-slit experiments, and one would like an analogous relation
for the case of 3-slit experiments. That is the focus of this paper.
Of late there has been a newly generated focus on the three-slit interference
experiments \cite{urbasi,hess,sam}, albeit for a different reason.

\section{Three-slit interference}

Three slit interference is somewhat more involved than its 2-slit counterpart
simply due to the fact that while the two-slit interference is the result
of interference
between two parts coming from the two slits, in the 3-slit interference
there are three parts which interfere in different ways. In the general case,
if we assume that the separation between slits 1 and 2 is $\ell_1$ and that
between slits 2 and 3 is $\ell_2$, there are two interferences from slit 1 and 2
and from slit 2 and slit 3. 
In addition there is an interference between parts from slit 1 and 3,
which involves a slit separation of $\ell_1+\ell_2$. In the case where the two
slit separations are the same, $\ell_1=\ell_2=d$, there are two interferences with
slit separation $d$ and one interference with slit separation $2d$.
Having more than two slits also allows, in principle, the possibility of having
different geometrical arrangement of slits. However, we restrict ourselves
to the case of three slits in a linear geometry, as shown in
Fig. \ref{trislit}, as that is the geometry in which the experiment is
usually done, and that is also the geometry which is used in previous
investigations of multi-slit experiments.

We expect additional complicacy in interpreting Bohr's complementarity because
if we know that the particle did not go through (say) slit 3, it may not
imply complete loss of interference as there is still ambiguity regarding
which of the other two slits, 1 or 2, the particle went through.

\subsection{Which-way information}

First we would like to have a way of knowing which of the three slits the
particle passed through. Any which-path detector should have three states
which should correlate with the particle passing through each slit.
Let these states be $|d_1\rangle, |d_2\rangle, |d_3\rangle$, which
correspond to particle passing through slits 1, 2 and 3, respectively.
Without loss of generality we assume that the states 
$|d_1\rangle, |d_2\rangle, |d_3\rangle$ are normalized, although they may
not necessarily be mutually orthogonal. The combined state of the particle
and the which-path detector can be written as
\begin{equation}
|\Psi\rangle = \sqrt{p_1}|\psi_1\rangle|d_1\rangle + \sqrt{p_2}|\psi_2\rangle|d_2\rangle
+\sqrt{p_3}|\psi_3\rangle|d_3\rangle,
\label{correlated}
\end{equation}
where $\sqrt{p_1}|\psi_1\rangle,\sqrt{p_2}|\psi_2\rangle,\sqrt{p_3}|\psi_3\rangle$ are the amplitudes
of the particle passing through the slit 1, 2 and 3, respectively.
Particle passes through slits 1, 2, and 3 with probabilities $p_1$, $p_2$
and $p_3$, respectively.

If $|d_1\rangle, |d_2\rangle, |d_3\rangle$ are mutually orthogonal, we
can find a Hermitian operator (and thus, a measurable quantity) which will
give us different eigenvalues corresponding to $|d_1\rangle, |d_2\rangle,
|d_3\rangle$, and thus to the particle passing through
each of the three slit. In this case, which slit the particle went through,
can be known without ambiguity.

If $|d_1\rangle, |d_2\rangle, |d_3\rangle$ are not mutually orthogonal, 
the three ways of particle going through the slits will not be fully
distinguishable. One needs to define a {\em distinguishability} of the three
different paths of the particle. Defining distinguishability in multi-slit
experiments has been a thorny issue \cite{jaeger,durr,bimonte,bimonte1,englertmb}.

\subsection{Unambiguous Quantum State Discrimination}

As one  can see from (\ref{correlated}), the problem of distinguishing
the three paths of the particle boils down to distinguishing between the
three states $|d_1\rangle, |d_2\rangle, |d_3\rangle$. In the following
we describe a well established method of {\em unambiguously} discriminating
between two non-orthogonal quantum states, which goes by the name of
Unambiguous Quantum State Discrimination
(UQSD) \cite{uqsd,dieks,peres,jaeger2,bergou}.
If two states $|p\rangle,|q\rangle$
are not orthogonal, it is impossible to distinguish between the two with
certainty. What UQSD does is to separate the measurement results into two
categories.  First category is the one in which the discrimination fails.
The second one distinguishes between the two states {\em without any error}.
Let the first system, whose states are $|p\rangle,|q\rangle$ interact with
a second system which is initially in a state $|s_0\rangle$. The interaction
and time evolution leads to the following entangled states
\begin{eqnarray}
\op{U}|p\rangle|s_0\rangle &=& \alpha|p_1\rangle|s_1\rangle + \beta|p_2\rangle|s_2\rangle\nonumber\\
\op{U}|q\rangle|s_0\rangle &=& \gamma|q_1\rangle|s_1\rangle + \delta|q_2\rangle|s_2\rangle,
\end{eqnarray}
where $\op{U}$ is a unitary operator such that $\langle s_1|s_2\rangle=0$
and $\langle p_1|q_1\rangle=0$. If one measures an observable of the second
system which has two eigenstates $|s_1\rangle,|s_2\rangle$ with different
eigenvalues,  the result $|s_1\rangle$ lands us into a situation where
the non-orthogonal states $|p\rangle,|q\rangle$ have been replaced by
orthogonal $|p_1\rangle,|q_1\rangle$. The orthogonal states
$|p_1\rangle,|q_1\rangle$ can be distinguished with hundred percent accuracy,
thus distinguishing the original $|p\rangle,|q\rangle$ without error.
However, the other result for system 2, $|s_2\rangle$, leads us to states
$|p_2\rangle,|q_2\rangle$ which are not orthogonal, and the discrimination
of $|p\rangle,|q\rangle$ fails. So, either the process fails, or it
distinguishes between $|p\rangle,|q\rangle$ without error. The probability
of successfully distinguishing between $|p\rangle$ and $|q\rangle$ depends
on the constants $\alpha,\beta,\gamma,\delta$. It can be easily shown that
the maximum probability of successfully distinguishing between
$|p\rangle$ and $|q\rangle$ is given by \cite{dieks}
\begin{equation}
P = 1 - |\langle p|q\rangle|.
\end{equation}
This is called the IDP (Ivanovic-Dieks-Peres) limit, and $|p\rangle$
and $|q\rangle$ cannot be distinguished unambiguously with a probability
larger than this, even in principle \cite{uqsd,dieks,peres}. 
In other words, UQSD is the best bet for discriminating between two
non-orthogonal states. This fact has also been experimentally demonstrated
recently \cite{Waldherr,Agnew}. Thus, the IDP limit sets a fundamental
bound on the distinguishability of two non-orthogonal states.

\subsection{Distinguishability}

The preceding analysis suggests a natural definition of path-distinguishability.
In a two-slit experiment, if the combined state of the particle and the
which way detector can be written as
$|\Psi\rangle = |\psi_1\rangle|d_1\rangle + |\psi_2\rangle|d_2\rangle$,
the two paths can be distinguished if the two states $|d_1\rangle,|d_2\rangle$
can be distinguished. The probability of successfully telling which slit
the particle went through is just the probability with which 
$|d_1\rangle,|d_2\rangle$ can be unambiguously distinguished.  We thus
define distinguishability of the two paths in a double-slit experiment
as the upper limit of the probability of unambiguously distinguishing between
$|d_1\rangle$ and $|d_2\rangle$. Thus we define a new path distinguishability
for a two-slit experiment as
\begin{equation}
{\mathcal D_Q} \equiv 1 - |\langle d_1|d_2\rangle|.
\label{D2}
\end{equation}
We denote it by ${\mathcal D_Q}$ to distinguish it from the ${\mathcal D}$
used in (\ref{egy}).

In the 3-slit experiment, the entangled state (\ref{correlated}) implies that,
after the particle has passed through the triple slit, the which-path
detector could be in state $|d_1\rangle$ or $|d_2\rangle$ or $|d_3\rangle$,
with probabilities  $p_1$, $p_2$ and $p_3$, respectively.
The probability with
which one can tell which of the three slits the particle went through,
is just the probability with which one can distinguish between 
$|d_1\rangle,|d_2\rangle,|d_3\rangle$. So, the problem of distinguishing
between the three paths boils down to distinguishing between three
non-orthogonal states $|d_1\rangle$, $|d_2\rangle$, $|d_3\rangle$, which
occur with different probabilities.

UQSD has also been generalized to the case of N non-orthogonal states.
The probability
of unambiguously distinguishing between N non-orthogonal quantum states is
bounded by \cite{zhang,qiu}
\begin{equation}
P_N \le 1 - {1\over N-1}\sum_{i\neq j} \sqrt{p_ip_j} |\langle d_i|d_j\rangle|,
\label{pn}
\end{equation}
where $\{|d_i\rangle\}$ are the $N$ non-orthogonal states, and $p_i$ are
their respective a-priori probabilities. For three non-orthogonal states,
the above reduces to
\begin{equation}
P_3 \le 1 - (\sqrt{p_1p_2}|\langle d_1|d_2\rangle|+
\sqrt{p_2p_3}|\langle d_2|d_3\rangle|+\sqrt{p_1p_3}|\langle d_1|d_3\rangle|).
\label{p3}
\end{equation}

We define the path-distinguishability
for the 3-slit experiment as the upper limit of the probability with which one can
distinguish between the three states $|d_1\rangle,|d_2\rangle,|d_3\rangle$,
which is now given by
\begin{equation}
{\mathcal D_Q} \equiv 1 - (\sqrt{p_1p_2}|\langle d_1|d_2\rangle| + 
\sqrt{p_2p_3}|\langle d_2|d_3\rangle| +\sqrt{p_1p_3}|\langle d_1|d_3\rangle|),
\label{D3}
\end{equation}
and whose value lies in the range $0\le {\mathcal D_Q}\le 1$. 
Eqn. (\ref{D3}) should be seen as a generalization of the IDP limit to
the case of three non-orthogonal states.

\subsection{Interference and which-way information}
\label{interference}

In order to obtain a tight bound on the visibility of interference, given a
particular amount of which-way information, we do a wave-packet
analysis in the following way.
We consider a particle traveling along the z-direction and passing
through the triple-slit, with a slit separations $\ell_1$ and $\ell_2$, and then
interacting with a which-path detector through a unitary evolution.
The evolution of the state is given by the time-dependent Schr\"{o}dinger equation,
\begin{equation}
i\hbar{\partial _{t}|{\Psi}(t)}\rangle =\mathcal{H}|\Psi(t)\rangle , 
\label{SE}
\end{equation}
where $ \mathcal{H} $ is the Hamiltonian of the system (the energy operator).

Our strategy is the following. The motion of the particle along the z-axis
is redundant, as it only translates the position of the particle from the
triple-slit to the screen. What is relevant is the motion and dispersion of the 
particle along the x-direction. Without going into the details of the which-path
detector, we assume that it is a device having 3 states $|d_1\rangle,
|d_2\rangle, |d_3\rangle$, which get entangled with the three amplitudes of
the particle passing through slits 1, 2 and 3, respectively. This entanglement
is a necessary condition for the which-path detector to get any information
about which slit the particle went through \cite{tqrv}.

In the following we assume that the mass of the particle is $m$, and the
three slits are located at $x = +\ell_1,~0,~-\ell_2$. The wavefunction emerging from a
particular slit is assumed to be a Gaussian wave-packet in the x-direction,
of width $\epsilon$, localized at $x = \ell_1,~0,~\text{or } -\ell_2$. A wavefunction of the form
$e^{-{(x-\ell_1)^2\over 4\epsilon^2}}$ should represent the particle emerging
from a slit of width  $\epsilon$, located at $x = \ell_1$, in a reasonable
approximation. As we shall see later, the choice of the Gaussian form does not
lead to any loss of generality because the width of the Gaussians does
not appear in the final results.

Using this strategy, the combined state of the particle and the which-path
detector, when the particle emerges from the triple-slit (time $t=0$),
will be of the form
\begin{eqnarray}
\Psi(x,0) = A \left(\sqrt{p_1}|d_1\rangle e^{-{(x-\ell_1)^2\over 4\epsilon^2}}
+\sqrt{p_2}|d_2\rangle e^{-{x^2\over 4\epsilon^2}}\right.
\left.+ \sqrt{p_3}|d_3\rangle e^{-{(x+\ell_2)^2\over 4\epsilon^2}}\right),
\label{initial}
\end{eqnarray}
where $A = (2\pi\epsilon^2)^{-1/4}$, and $p_1+p_2+p_3=1$. It represents three
Gaussian wave-packets localised at the three slits,
namely $x=\ell_1$, $x=0$ and $x=-\ell_2$, entangled with the
three states of the which-path detector. The width of the three Gaussians
is chosen to be the same because the widths of the three slits is assumed
to be the same. It is not difficult to see, from the subsequent analysis,
that even if we consider the three slit-widths to be different, e.g.,
$\epsilon_1,~ \epsilon_2,~\epsilon_3$, they would all drop out from our
final result.  In general one should consider
a factor $\sqrt{p_1}e^{i\theta}$ instead of $\sqrt{p_1}$, and likewise
for $\sqrt{p_2}$ and $\sqrt{p_3}$.
We do not consider these phase factors simply 
because they will be absorbed in certain phases introduced later, in
section {\ref{interference}}.
We consider the probabilities of all the three paths to be different for
generality, but point out that unequal probabilities reduce the
visibility of interference.  Even in the two-slit interference, unequal
beams reduce the visibility of interference.

After a time $t$, the state of the particle and the detector evolves
via (\ref{SE}), with $ \mathcal{H} = \op{p}_x^2/2m$, to
\begin{eqnarray}
\Psi(x,t) &=& A \left(\sqrt{p_1}|d_1\rangle e^{-{(x-\ell_1)^2\over 4\epsilon^2+2i\hbar t/m}}
+\sqrt{p_2}|d_2\rangle e^{-{x^2\over 4\epsilon^2+2i\hbar t/m}}\right.
\left.+ \sqrt{p_3}|d_3\rangle e^{-{(x+\ell_2)^2\over 4\epsilon^2+2i\hbar t/m}}\right),
\end{eqnarray}
where $A = [\sqrt{2\pi}(\epsilon+i\hbar t/2m\epsilon)]^{-1/2}$.
It represents the three wave-packets spreading and overlapping with each
other. The smaller the width $\epsilon$ of the slits, the stronger is the
overlap between the three wave-packets.

The probability (density) of finding the particle at a position $x$ on
the screen is given by
\begin{eqnarray}
|\Psi(x,t)|^2 &=& |A|^2 \left(p_1 e^{-{(x-\ell_1)^2\over 2\sigma^2}} + p_2 e^{-{x^2\over 2\sigma^2}}
+ p_3 e^{-{(x+\ell_2)^2\over 2\sigma^2}}\right.\nonumber\\
&&+ \sqrt{p_1p_2}e^{-{2x^2+\ell_1^2-2x\ell_1\over 4\sigma^2}}\left[
\langle d_1|d_2\rangle  e^{{ix\ell_1\hbar t/m -i\hbar t\ell_1^2/2m \over 4\Omega^2}}
+ \langle d_2|d_1\rangle  
e^{-{ix\ell_1\hbar t/m -i\hbar t\ell_1^2/2m \over 4\Omega^2}}\right]\nonumber\\
&&+ \sqrt{p_2p_3}e^{-{2x^2+\ell_2^2+2x\ell_2\over 4\sigma^2}}\left[\langle d_2|d_3\rangle  
e^{{ix\ell_2\hbar t/m +i\hbar t\ell_2^2/2m \over 4\Omega^2}}
+\langle d_3|d_2\rangle  
e^{-{ix\ell_2\hbar t/m +i\hbar t\ell_2^2/2m\over 4\Omega^2}}\right] \nonumber\\
&&+ \sqrt{p_1p_3}e^{-{x^2+(\ell_1^2+\ell_2^2)/2+x(\ell_2-\ell_1)\over 2\sigma^2}}\left[
\langle d_1|d_3\rangle  e^{{i[{x(\ell_1+\ell_2)}+{(\ell_2^2-\ell_1^2)/2}]{\hbar t/ m}\over 4\Omega^2}}\right.\nonumber\\
&&\left.\left.+ \langle d_3|d_1\rangle  
e^{-{i[{x(\ell_1+\ell_2)}+{(\ell_2^2-\ell_1^2)/2}]{\hbar t/m}\over 4\Omega^2}}\right]\right),
\end{eqnarray}
where $\sigma^2 = \epsilon^2 + (\hbar t/2m\epsilon)^2$ and
$\Omega^2 = \epsilon^4 + (\hbar t/2m)^2$.
We write the overlaps of various detector states as:
$\langle d_1|d_2\rangle = |\langle d_1|d_2\rangle|e^{i\theta_1}$,
$\langle d_2|d_3\rangle = |\langle d_3|d_3\rangle|e^{i\theta_2}$,
$\langle d_1|d_3\rangle = |\langle d_1|d_3\rangle|e^{i\theta_3}$.
The phases $\theta_1, \theta_2, \theta_3$ being arbitrary, any additional
phases coming from the amplitudes of the initial state (\ref{initial}) 
would be absorbed in them.  The probability density then reduces to
\begin{eqnarray}
|\Psi(x,t)|^2 &=& |A|^2 \left(e^{-{x^2\over 2\sigma^2}} \left (p_1 e^{-{\ell_1^2-2x\ell_1\over 2\sigma^2}} + p_2
+ p_3 e^{-{\ell_2^2+2x\ell_2\over 2\sigma^2}}\right)\right.\nonumber\\
&& + 2\sqrt{p_1p_2}|\langle d_1|d_2\rangle |  e^{-{2x^2+\ell_1^2-2x\ell_1\over 4\sigma^2}}
\cos\left( {{x\ell_1\hbar t\over 4m\Omega^2} - \beta_1 + \theta_1}\right) \nonumber\\
&&+ 2 \sqrt{p_2p_3}|\langle d_2|d_3\rangle  |  e^{-{2x^2+\ell_2^2+2x\ell_2\over 4\sigma^2}}
\cos \left({{x\ell_2\hbar t\over 4m\Omega^2} + \beta_2 + \theta_2} \right) \nonumber\\
&& +\left.  2\sqrt{p_1p_3}|\langle d_1|d_3\rangle |  e^{-{2x^2+\ell_1^2+\ell_2^2+2x(\ell_2-\ell_1)\over 4\sigma^2}}
{\cos \left( {x(\ell_1+\ell_2)\hbar t\over 4m\Omega^2} + \beta_3 + \theta_3 \right)}\right), 
\label{pattern}
\end{eqnarray}
where $\beta_1 = {\hbar t\ell_1^2\over 8m\Omega^2}$,
$\beta_2 = {\hbar t\ell_2^2\over 8m\Omega^2}$,
and $\beta_3 = {\hbar t(\ell_2^2-\ell_1^2)\over 8m\Omega^2}$.

Visibility of the interference fringes is conventionally defined as \cite{born}
\begin{equation}
{\mathcal V} = {I_{max} - I_{min} \over I_{max} + I_{min} } ,
\end{equation}
where $I_{max}$ and $I_{min}$ represent the maximum and minimum intensity
in neighbouring fringes, respectively. Since all the factors multiplying
the cosine terms in (\ref{pattern}) are non-negative, the maxima of the
fringe pattern will occur where the value of all the cosines is $+1$.
The minima will occur where value of all the cosines is $-1/2$
at the same time.

\subsection{Equally spaced slits}

In the usual three-slit interference experiments, the three slits are
equally spaced. In that case $\ell_1 = \ell_2 \equiv d$, and
$\beta_1=\beta_2 \equiv \beta$ and $\beta_3=0$.
Provided we ignore $\beta$, the cosines can indeed have values all $+1$ 
or all $-1/2$ for certain values of $x$.
Ignoring $\beta$ amounts to ignoring $d/2$ in comparison to $x$, which is
justified if one looks at any fringe except the central one (at $x=0$),
since fringe width on the screen is much larger than the slit separation $d$.
The {\em ideal} visibility can then be written down as
\begin{equation}
{\mathcal V}_I = {3\left( \sqrt{p_1p_2}|\langle d_1|d_2\rangle| e^{{xd\over 2\sigma^2}} + \sqrt{p_1p_3}|\langle d_1|d_3\rangle| e^{-{d^2\over 2\sigma^2}} + \sqrt{p_2p_3}|\langle d_2|d_3\rangle| e^{-{xd\over 2\sigma^2}}\right) \over \alpha +\sqrt{p_1p_2}|\langle d_1|d_2\rangle| e^{{xd\over 2\sigma^2}} + \sqrt{p_1p_3}|\langle d_1|d_3\rangle| e^{-{d^2\over 2\sigma^2}} + \sqrt{p_2p_3}|\langle d_2|d_3\rangle| e^{-{xd\over 2\sigma^2}} }.
\end{equation}
where $\alpha = 2 \left(p_2 + [p_1e^{x d \over \sigma^2} + p_3e^{-x d \over \sigma^2}]e^{-{d^2\over 2\sigma^2}}\right)$.
In reality, fringe visibility will be reduced due many factors,
including the width of the slits. For example, if
the width of the slits is very large, the fringes may not be visible at all.
It will also get reduced by the varying phases $\theta_1,\theta_2,\theta_3$.
In an ideal situation, the maximum visibility one can theoretically get will
be in the case when $d \ll \sigma$, which amounts to assuming that the spread
of a particular wave-packet, when it reaches the screen, is so large
that the separation between two neighboring slits $d$ is negligible in
comparison, and also when all
the phases are either zero or their effect cancels out. Actual fringe visibility will be
less than or equal to that, and can be written as
\begin{equation}
{\mathcal V} \le {3\left( \sqrt{p_1p_2}|\langle d_1|d_2\rangle| + \sqrt{p_1p_3}|\langle d_1|d_3\rangle| + \sqrt{p_2p_3}|\langle d_2|d_3\rangle| \right) \over 2 +\sqrt{p_1p_2}|\langle d_1|d_2\rangle| + \sqrt{p_1p_3}|\langle d_1|d_3\rangle| + \sqrt{p_2p_3}|\langle d_2|d_3\rangle| }.
\label{v3}
\end{equation}
Using (\ref{D3}) the above equation gives 
\begin{equation}
{\mathcal V} + {2{\mathcal D_Q}\over 3- {\mathcal D_Q}} \le 1 .
\label{duality}
\end{equation}
Eqn. (\ref{duality}) is a new duality relation which puts a bound on how
much which-way information we can obtain and how much fringe visibility we
can get at the same time. It is straightforward to check that
${\mathcal V}=1$ is
possible only for ${\mathcal D_Q}=0$, and ${\mathcal D_Q}=1$ implies 
${\mathcal V}=0$. Note that (\ref{duality}) can also be expressed in 
another form
\begin{equation}
{\mathcal D_Q} + {2{\mathcal V}\over 3- {\mathcal V}} \le 1 .
\end{equation}

\subsection{Unequally spaced slits}

When the three slits are equally spaced, maximum is achieved when
the two $\cos\left({xd\hbar t\over 4m\Omega^2}\right)$ terms and the
$\cos\left({2xd\hbar t\over 4m\Omega^2}\right)$ term have values +1
at the same time, which happens when ${xd\hbar t\over 4m\Omega^2}=2n\pi$,
provided that $\theta_1,\theta_2,\theta_3$ are zero.
When the three slits are unequal, the three terms 
$\cos\left({x\ell_1\hbar t\over 4m\Omega^2}\right)$, $\cos\left({x\ell_2\hbar t\over 4m\Omega^2}\right)$ and
$\cos\left({x(\ell_1+\ell_2)\hbar t\over 4m\Omega^2}\right)$ cannot all have
values +1 at the same time. Thus the maximum intensity will be smaller
than that in the equally spaced case, or
$I_{max}^{unequal} < I_{max}^{equal}$.

Minimum intensity in the equally spaced case is attained when all the
cosine terms are equal to $-1/2$, which happens when
${xd\hbar t\over 4m\Omega^2}={2n\pi\over 3}$, provided that $\theta_1,\theta_2,\theta_3$ are zero.
When the three slits are unequally spaced, the three terms
$\cos\left({x\ell_1\hbar t\over 4m\Omega^2}\right)$, $\cos\left({x\ell_2\hbar t\over 4m\Omega^2}\right)$ and
$\cos\left({x(\ell_1+\ell_2)\hbar t\over 4m\Omega^2}\right)$ cannot all have
values $-1/2$ at the same time. Thus the minimum intensity will be larger
than that in the equally spaced case, or
$I_{min}^{unequal} > I_{min}^{equal}$. These two observations lead to
the straightforward conclusion that fringe visibility in the case of 
unequal slits will be strictly smaller than that in the case of equal slits,
other things being the same,
\begin{equation}
{\mathcal V}^{unequal} < {\mathcal V}^{equal}.
\end{equation}
Using the above in conjunction with (\ref{v3}) we can write for the
fringe visibility, for the case of unequal slits
\begin{equation}
{\mathcal V} < {3\left( \sqrt{p_1p_2}|\langle d_1|d_2\rangle| + \sqrt{p_1p_3}|\langle d_1|d_3\rangle| + \sqrt{p_2p_3}|\langle d_2|d_3\rangle| \right) \over 2 +\sqrt{p_1p_2}|\langle d_1|d_2\rangle| + \sqrt{p_1p_3}|\langle d_1|d_3\rangle| + \sqrt{p_2p_3}|\langle d_2|d_3\rangle| }.
\label{v3u}
\end{equation}
Using (\ref{D3}) the above equation gives 
\begin{equation}
{\mathcal V} + {2{\mathcal D_Q}\over 3- {\mathcal D_Q}} < 1 .
\label{dualu}
\end{equation}
Note the strictly less than sign in the above, in contrast with (\ref{duality}).
The above relation implies that if the slits are unequally spaced, even if
the distinguishability ${\mathcal D}_Q$ is reduced to zero, the visibility
of interference can never be 1. The reason of this behaviour lies in the
fact that a certain amount of loss of interference visibility is rooted
not in the path distinguishability, but in the unequal spacing of the
three slits.

\subsection{Specific cases}

Let us look at some special cases arising from the fact that there are not
two, but three slit. Suppose we have a which-path detector which can
detect with certainty if the particle has passed through slit 1 or not.
If the particle has not passed through slit 1, the detector is unable to say
which of the other two slits, 2 or 3, has the particle taken. Such a
scenario can occur, for example, if we have a tiny camera in front of
slit 1 which, for each particle, can say if the particle has gone through
slit 1 or not.
In this case $|d_1\rangle$ is orthogonal to both $|d_2\rangle$ and $|d_3\rangle$,
and $|d_2\rangle$, $|d_3\rangle$ are parallel. Thus, in this case,
$|\langle d_2|d_3\rangle|=1$ and
$|\langle d_1|d_2\rangle|=|\langle d_1|d_3\rangle|=0$. Assuming, for simplicity,
$p_1=p_2=p_3={1\over 3}$, the distinguishability ${\mathcal D_Q}$, in this case is
\begin{equation}
{\mathcal D_Q} \equiv 1 - {1\over 3}(|\langle d_1|d_2\rangle| + 
|\langle d_2|d_3\rangle| +|\langle d_1|d_3\rangle|) = {2\over 3},
\end{equation}
Consequently, the fringe visibility is limited by
\begin{equation}
{\mathcal V} \le {3\over 7}.
\end{equation}

Physically what is happening is the following. Particle going through slits
2 and 3 gives rise to a sharp interference pattern, however, particle going
through slit 1 gives rise to a uniform background particle count, thus
reducing the overall visibility of the fringes arising from slits 2 and 3.

Let us consider another case where $|d_1\rangle$ and $|d_2\rangle$ are 
orthogonal to each other, but both have equal overlap with $|d_3\rangle$.
Such a case can be exemplified by
$|d_1\rangle={1\over\sqrt{2}}(|\uparrow\rangle+|\downarrow\rangle)$,
$|d_2\rangle={1\over\sqrt{2}}(|\uparrow\rangle-|\downarrow\rangle)$ and
$|d_3\rangle=|\uparrow\rangle$, where $|\uparrow\rangle,~|\downarrow\rangle$
form an orthonormal set. In this case (again, for simplicity, assuming
$p_1=p_2=p_3={1\over 3}$) the distinguishability is
\begin{equation}
{\mathcal D_Q} = 1 - {1\over 3}(|\langle d_1|d_2\rangle| + 
|\langle d_2|d_3\rangle| +|\langle d_1|d_3\rangle|) = 1-{\sqrt{2}\over 3},
\end{equation}
Consequently, the fringe visibility is limited by
\begin{equation}
{\mathcal V} \le {3\sqrt{2}\over 6+\sqrt{2}}.
\end{equation}

\subsection{The two-slit experiment}

Just for completeness, here we wish to derive a duality relation for
a two-slit experiment in the case where one defines path
distinguishability based on UQSD. We define the distinguishability of two
paths as the upper limit of (\ref{pn}) for N=2. The path distinguishability
then reads
\begin{equation}
{\mathcal D_Q} \equiv 1 - 2\sqrt{p_1p_2}|\langle d_1|d_2\rangle|,
\label{D2E}
\end{equation}
where $p_1$ and $p_2$ are the probabilities of the particle to go
through the first and the second slit, respectively.

By carrying out an analysis similar to the one earlier in this section
(essentially, putting $p_3=0$ in (\ref{pattern}) and redefining some
constants),
one can show that the distinguishability and fringe visibility, in the
two-slit experiment, are bounded by
\begin{equation}
{\mathcal V}  + {\mathcal D_Q} \le 1 .
\label{Dnew}
\end{equation}
The  above relation is a simple wave-particle duality relation for a
two-slit interference experiment where the beams may be unequal.

It can be connected to Englert's duality relation (\ref{egy}) for the case
$p_1 = p_2=1/2$. In Englert's analysis, distinguishability is given by
${\mathcal D} \equiv \sqrt{1 - |\langle d_1|d_2\rangle|^2}$ \cite{englert},
which can be related to ${\mathcal D_Q}$ in (\ref{D2E}), 
for $p_1 = p_2=1/2$,  by the relation
\begin{equation}
{\mathcal D_Q}  = 1 -  \sqrt{ 1 - {\mathcal D}^2} .
\end{equation}
If one plugs in the above form of ${\mathcal D_Q}$ in (\ref{Dnew}),
the latter reduces to
${\mathcal V}^2 + {\mathcal D}^2 \le 1$, which is just 
Englert's duality relation (\ref{egy}). The new
relation (\ref{Dnew}) appears to be more versatile for two-slit experiments,
because it also applies
to certain modified two-slit experiments in which the which-path detector
is replaced by a ``quantum device" \cite{tqduality}.

Lastly we discuss a particular scenario in which the two states of the
which-path detector are identical, namely $|\langle d_1|d_2\rangle| = 1$.
In such a situation, experimentally one cannot tell which slit the
particle went through. However, if the probabilities of the particle
for going through the two slits are known to be different, one
can {\em predict} which slit the particle is most likely to have gone
through. In this situation our ${\mathcal D_Q}$, given by (\ref{D2E}),
is reduced to
\begin{equation}
\overline{\mathcal D}_Q \equiv 1 - 2\sqrt{p_1p_2}.
\end{equation}
Interestingly the above reduced distinguishability is related to the
{\em predictability}, defined by Greenberger and
Yasin \cite{greenberger} as ${\mathcal P} \equiv |p_1 - p_2|$,
by the following relation
\begin{equation}
\overline{\mathcal D}_Q = 1 - \sqrt{1 - {\mathcal P}^2}.
\end{equation}
So, for the case  $|\langle d_1|d_2\rangle| = 1$, our new duality relation
(\ref{Dnew}) reduces to
\begin{equation}
{\mathcal P}^2 + {\mathcal V}^2 \le 1 ,
\label{greenberger}
\end{equation}
which is precisely the duality relation derived by Greenberger and
Yasin \cite{greenberger}.

So, the versatility of the new two-slit duality relation can be seen from the
fact that for $p_1=p_2$ it reduces to Englert's duality relation dealing
with {\em distinguishability}, and for $|\langle d_1|d_2\rangle| = 1$,
it reduces
to Greenberger and Yasin's duality relation dealing with {\em predictability}.

\section{Conclusion}

In the analysis carried out in this paper, we have introduced
a new path distinguishability ${\mathcal D_Q}$, based on UQSD, which 
is just the upper limit of the 
probability with which one can {\em unambiguously} distinguish between 
the quantum states of the which-path detector correlated with the paths
of the particle. Consequently, it is the maximum probability with which
one can {\em unambiguously} tell which slit the particle went through.
We carried out a wave-packet evolution of a particle through a
triple-slit. Calculating the fringe-visibility after a Schr\"odinger evolution,
we relate it to the path distinguishability and derive a new duality
relation ${\mathcal V} + {2{\mathcal D_Q}\over 3- {\mathcal D_Q}} \le 1$.
The analysis is restricted to three slits
of equal widths, in a linear geometry, as shown in Fig. \ref{trislit}.
Starting from the triple-slit, the time evolution, leading to the probability
density of the particle on the screen, is exact. Various approximations, 
in the subsequent analysis, are made only to obtain the maximum possible
visibility of interference, given a particular ${\mathcal D_Q}$.
Because of the way in which the analysis is carried out, this should be
the tightest possible bound on distinguishability and fringe visibility
for the 3-slit experiment.
For two-slit interference, we derive a new duality relation which reduces
to Englert's duality relation and Greenberger and Yasin's duality relation,
in different limits.
Lastly, we feel that (\ref{pn}) suggests
a straightforward definition of distinguishability
for N-slit interference experiments:
\begin{equation}
{\mathcal D_Q} = 1 - {1\over N-1}\sum_{i\neq j} \sqrt{p_ip_j} |\langle d_i|d_j\rangle|,
\label{DN}
\end{equation}
where $|d_i\rangle$ is the state of the path-detector correlated with 
the i'th of the N possible paths.

\section*{Acknowledgement}

Mohd Asad Siddiqui thanks the University Grants Commission, India for 
financial support.

\end{document}